\documentclass[12pt]{article}
\usepackage{epsfig}
\begin{document}
\title{Photoproduction of $\eta^\prime$ mesons off nuclei \\
 and their properties in the nuclear medium}
\author{E. Ya. Paryev\\
{\it Institute for Nuclear Research, Russian Academy of Sciences,}\\
{\it Moscow 117312, Russia}}

\renewcommand{\today}{}
\maketitle

\begin{abstract}
   We study the photoproduction of $\eta^\prime$ mesons from nuclei near the threshold within
   the collision model based on the nuclear spectral function. The model takes properly into
   account both primary photon--nucleon ${\gamma}N \to \eta^\prime$N and secondary
   pion--nucleon ${\pi}N \to {\eta^\prime}N$ production processes as well as the effect of the
   nuclear $\eta^\prime$ mean-field potential on these processes. We find that the secondary
   channel ${\pi}N \to {\eta^\prime}N$ plays an insignificant role in the $\eta^\prime$ photoproduction
   off nuclei. We calculate the transparency ratio for $\eta^\prime$ mesons and compare it to the
   available experimental data. The comparison indicates an inelastic ${\eta^\prime}N$ cross section
   of the order of 6--10 mb. We show that the existing transparency ratio measurements do not allow one
   to distinguish between two adopted $\eta^\prime$ in-medium modification scenarios.
   Our studies also demonstrate that the momentum distribution and excitation function for $\eta^\prime$
   production in ${\gamma}A$ reactions reveal some sensitivity to these scenarios, which means that such
   observables may be an important tool to get a valuable information on the $\eta^\prime$ in-medium
   properties.

\end{abstract}

\newpage

\section*{1. Introduction}

\hspace{1.5cm} The study of the modifications of the hadronic properties (masses and decay widths) in
a strongly interacting environment is one of the most important subjects in contemporary
hadron and nuclear physics owing to the expectation to observe a partial restoration of chiral symmetry
in nuclear medium.
The production of mostly light vector mesons $\rho$, $\omega$, $\phi$ in nuclear reactions with
elementary probes (photons, protons) as well as in heavy--ion collisions has been investigated
to search for possible in-medium modifications of their properties (see, for example, [1--13]).

  Another interesting case of hadron medium renormalization
is that of the pseudoscalar ${\eta^\prime}(958)$ meson, whose mass, as is believed, is generated by the
interplay of the QCD $U_A(1)$ anomaly and chiral symmetry breaking. The $\eta^\prime$ mass reduction
induced by the combined effect of the density-independent $U_A(1)$ anomaly and
partial restoration of chiral symmetry in nuclear matter is predicted to be of the order of 150 MeV at the
saturation density $\rho_0$ [14, 15]. The possible suppression of the $U_A(1)$ anomaly effect in nuclear
medium leads to further mass reduction of the $\eta^\prime$ here, which is expected to be in the 100 MeV area
at the normal nuclear matter density [14, 15]. A reduction of the $\eta^\prime$ mass of 200 MeV in medium
formed in heavy--ion collisions was recently deduced from experiments at RHIC [16].
On the other hand, it is expected to be only
a few ten MeV at saturation density [15] if it is estimated in the linear density approximation by using
the ${\eta^\prime}N$ scattering length extracted from the study of the $pp \to pp{\eta^\prime}$ reaction near
the threshold at COSY [17]. The $\eta^\prime$ mass reduction in matter due to the possible suppression
of the $U_A(1)$ anomaly effect in it is not accompanied by the appearance of an additional inelastic
${\eta^\prime}N$ processes in the nuclear medium [14, 15]. Hence, the ${\eta^\prime}$ absorption can be
small, which is in line with the recent theoretical [18] and experimental [19] findings. Thus, the
${\eta^\prime}$ in-medium properties should provide us valuable information both on the
partial restoration of chiral symmetry at finite density and on the behavior
of the $U_A(1)$ anomaly in the nuclear medium.

   The production of $\eta^\prime$ mesons in photon collisions with C, Ca, Nb, and Pb targets has been studied
via the $\eta^\prime \to {\pi^0}{\pi^0}\eta \to 6\gamma$ decay channel by the CBELSA/TAPS Collaboration in [19].
The variation of the $\eta^\prime$ nuclear transparency ratio normalized to carbon with atomic number $A$ and
with $\eta^\prime$ momentum $p_{\eta^\prime}$ as well as the momentum distribution of $\eta^\prime$ mesons
produced off a C target for the incident photon energy range of $E_{\gamma}=$1.5--2.2 GeV have been measured.
A comparison of the data for the transparency ratio as a function of the nuclear mass number $A$ at photon energies
of 1.7, 1.9 and 2.1 GeV with the theoretical calculations assuming a single step process for  $\eta^\prime$
creation and a one-body  $\eta^\prime$ absorption process as well as describing the propagation of
 $\eta^\prime$ mesons in nuclei in the eikonal approximation, yielded a width of the  $\eta^\prime$ meson
 of the order of 15--25 MeV in the nuclear rest frame at normal nuclear matter density $\rho_0$ and for an average
 momentum $p_{\eta^\prime}=1.05$ GeV/c [19]. In the low-density approximation, this corresponds to an in-medium
${\eta^\prime}N$ inelastic cross section of $\sigma_{\rm {\eta^\prime}N}\approx$ 6--10 mb [19]. In an extreme case,
providing the lower boundary for the determination of $\sigma_{\rm {\eta^\prime}N}$, when the one- and two-body
$\eta^\prime$ absorption mechanisms contribute equally, the extracted inelastic cross section is reduced to
$\sigma_{\rm {\eta^\prime}N}\approx$ 3--5 mb [19]. This allowed one to estimate in [19] the inelastic
${\eta^\prime}N$ cross section as 3--10 mb to take into account the uncertainties arising from the unknown strength
of the two-nucleon ${\eta^\prime}$ absorption mechanism. The small ${\eta^\prime}$ in-medium width, deduced in [19],
implies the feasibility of an experimental observation of ${\eta^\prime}$ bound states in nuclei [14, 15, 20].
In order to get a deeper insight into the interaction of the ${\eta^\prime}$ meson with nuclear matter, it is
important to analyze not only the $A$, but also the momentum dependences of the transparency ratio and the
${\eta^\prime}$ production cross section measured in [19]--the main aim of the present work.

In this paper, we study the inclusive ${\eta^\prime}$ creation in photon--nucleus collisions on the basis
of the nuclear spectral function approach [21--24]. We present detailed predictions for the absolute and
relative  ${\eta^\prime}$ meson production cross sections from these collisions in the threshold energy
region obtained within this approach in two scenarios for its in-medium modification by considering
corresponding primary photon--nucleon and secondary pion--nucleon
\footnote{$^)$Not considered quantitatively in [19].}$^)$
 ${\eta^\prime}$ production processes as well as compare part of them with the available data [19].

\section*{2. Framework}
\section*{2.1. One-step  ${\eta^\prime}$ production mechanism}

\hspace{1.5cm} Since we are interested in the incident photon energy range up to approximately
2 GeV, we accounted for the following direct elementary processes,
which have the lowest free production threshold ($\approx$ 1.446 GeV):
\begin{equation}
\gamma+p \to \eta^\prime+p,
\end{equation}
\begin{equation}
\gamma+n \to \eta^\prime+n.
\end{equation}
We will not consider the ${\eta^\prime}$ creation processes with an additional pion in the final
state because their total cross section is less than those of the reactions (1) and (2) by a factor
of about 3 [25] in the energy domain of our interest. In the following calculations we will include
the medium modification of the ${\eta^\prime}$ mesons, participating in the production processes (1),
(2) by using, for reasons of simplicity, their average in-medium mass $<m^*_{{\eta^\prime}}>$ defined as:
\begin{equation}
<m^*_{{\eta^\prime}}>=\int d^3r{\rho_N({\bf r})}m^*_{{\eta^\prime}}({\bf r})/A,
\end{equation}
where ${\rho_N({\bf r})}$ and $m^*_{{\eta^\prime}}({\bf r})$ are the local nucleon density and
${\eta^\prime}$ effective mass inside the nucleus, respectively. Assuming in line with [14, 15] that
\begin{equation}
m^*_{{\eta^\prime}}({\bf r})=m_{{\eta^\prime}}+V_0\frac{{\rho_N({\bf r})}}{{\rho_0}},
\end{equation}
we can readily rewrite Eq. (3) in the form
\begin{equation}
<m^*_{{\eta^\prime}}>=m_{{\eta^\prime}}+V_0\frac{<{\rho_N}>}{{\rho_0}}.
\end{equation}
Here, $m_{{\eta^\prime}}$ is the ${\eta^\prime}$ free space mass and $<{\rho_N}>$ is the
average nucleon density. For the potential depth at saturation density
$V_0\approx-100$ MeV [14, 15, 26] as well as for $<{\rho_N}>={\rho_0}/2$ [27], Eq. (5)
leads to
\begin{equation}
<m^*_{{\eta^\prime}}>=m_{{\eta^\prime}}+\frac{1}{2}V_0\approx m_{{\eta^\prime}}-50~
\mbox{{\rm MeV}}.
\end{equation}
In order to see the sensitivity of the ${\eta^\prime}$ production cross sections from the one-step
processes (1), (2) to the ${\eta^\prime}$ mass shift (6)
\footnote{$^)$We will quote namely this mass shift in figures 7--12, presented below. However, it should
be mentioned that in the literature to quantify the hadron in-medium mass shift it is usually quoted at
normal nuclear matter density.}$^)$,
we will also ignore it in our calculations.
The total energy $E^\prime_{{\eta^\prime}}$ of the ${\eta^\prime}$ meson inside the nuclear medium can be
expressed through its average effective mass $<m^*_{{\eta^\prime}}>$ defined above and in-medium momentum
${\bf p}^{\prime}_{{\eta^\prime}}$ as in the free particle case, namely:
\begin{equation}
E^\prime_{{\eta^\prime}}=\sqrt{({\bf p}^{\prime}_{{\eta^\prime}})^2+(<m^*_{{\eta^\prime}}>)^2}.
\end{equation}
The momentum ${\bf p}^{\prime}_{{\eta^\prime}}$ is related to the vacuum one ${\bf p}_{{\eta^\prime}}$
by the following expression:
\begin{equation}
\sqrt{({\bf p}^{\prime}_{{\eta^\prime}})^2+(<m^*_{{\eta^\prime}}>)^2}=
\sqrt{{\bf p}^2_{{\eta^\prime}}+m^2_{{\eta^\prime}}}.
\end{equation}

  Finally, neglecting the distortion of the incident photon and describing the ${\eta^\prime}$ meson
final-state absorption by in-medium cross section $\sigma_{\rm {\eta^\prime}N}$ as well as using the
results given in [21--24], we can represent the inclusive differential and total
\footnote{$^)$In the full phase space without any cuts on angle and momentum of the observed ${\eta^\prime}$ meson.}$^)$
cross sections for the production of ${\eta^\prime}$ mesons with the momentum ${\bf p}_{{\eta^\prime}}$
off nuclei in the primary photon--induced reaction channels (1), (2) as follows:
\begin{equation}
\frac{d\sigma_{{\gamma}A\to {\eta^\prime}X}^{({\rm prim})}
(E_{\gamma})}
{d{\bf p}_{\eta^\prime}}=I_{V}[A]
\end{equation}
$$
\times
\left[\frac{Z}{A}\left<\frac{d\sigma_{{\gamma}p\to {\eta^\prime}p}({\bf p}_{\gamma},
{\bf p}^{\prime}_{\eta^\prime})}{d{\bf p}^{\prime}_{\eta^\prime}}\right>_A+
\frac{N}{A}\left<\frac{d\sigma_{{\gamma}n\to {\eta^\prime}n}({\bf p}_{\gamma},{\bf p}^{\prime}_{\eta^\prime})}{d{\bf p}^{\prime}_{\eta^\prime}}\right>_A\right]\frac{d{\bf p}^{\prime}_{{\eta^\prime}}}{d{\bf p}_{{\eta^\prime}}},
$$
\begin{equation}
\sigma_{{\gamma}A\to {\eta^\prime}X}^{({\rm prim})}
(E_{\gamma})=I_{V}[A]
\left[\frac{Z}{A}\left<\sigma_{{\gamma}p\to {\eta^\prime}p}({\bf p}_{\gamma})\right>_A+
\frac{N}{A}\left<\sigma_{{\gamma}n\to {\eta^\prime}n}({\bf p}_{\gamma})\right>_A\right];
\end{equation}
where
\begin{equation}
I_{V}[A]=2{\pi}A\int\limits_{0}^{R}r_{\bot}dr_{\bot}
\int\limits_{-\sqrt{R^2-r_{\bot}^2}}^{\sqrt{R^2-r_{\bot}^2}}dz
\rho(\sqrt{r_{\bot}^2+z^2})
\exp{\left[-\sigma_{{\eta^\prime}N}A\int\limits_{z}^{\sqrt{R^2-r_{\bot}^2}}
\rho(\sqrt{r_{\bot}^2+x^2})dx\right]},
\end{equation}
\begin{equation}
\left<\frac{d\sigma_{{\gamma}N\to {\eta^\prime}N}({\bf p}_{\gamma},
{\bf p}^{\prime}_{\eta^\prime})}
{d{\bf p}^{\prime}_{\eta^\prime}}\right>_A=
\int\int
P_A({\bf p}_t,E)d{\bf p}_tdE
\left[\frac{d\sigma_{{\gamma}N\to {\eta^\prime}N}(\sqrt{s},<m^*_{{\eta^\prime}}>,{\bf p}^{\prime}_{\eta^\prime})}
{d{\bf p}^{\prime}_{\eta^\prime}}\right],
\end{equation}
\begin{equation}
\left<\sigma_{{\gamma}N\to {\eta^\prime}N}({\bf p}_{\gamma})\right>_A=
\int\int
P_A({\bf p}_t,E)d{\bf p}_tdE
\sigma_{{\gamma}N\to {\eta^\prime}N}(\sqrt{s},<m^*_{{\eta^\prime}}>)
\end{equation}
and
\begin{equation}
  s=(E_{\gamma}+E_t)^2-({\bf p}_{\gamma}+{\bf p}_t)^2,
\end{equation}
\begin{equation}
   E_t=M_A-\sqrt{(-{\bf p}_t)^2+(M_{A}-m_{N}+E)^{2}}.
\end{equation}
Here,
$d\sigma_{{\gamma}N\to {\eta^\prime}N}(\sqrt{s},<m^*_{{\eta^\prime}}>,{\bf p}^{\prime}_{\eta^\prime}) /d{\bf p}^{\prime}_{\eta^\prime}$ and $\sigma_{{\gamma}N\to {\eta^\prime}N}(\sqrt{s},<m^*_{{\eta^\prime}}>)$
are, respectively, the off-shell differential and total cross sections for the production of ${\eta^\prime}$
with reduced mass $<m^*_{{\eta^\prime}}>$ in reactions (1) and (2)
at the ${\gamma}N$ center-of-mass energy $\sqrt{s}$; ${\bf p}_{\gamma}$ is the momentum of the initial photon;
$\rho({\bf r})$ and $P_A({\bf p}_t,E)$ are the local nucleon density and the
spectral function
\footnote{$^)$Which represents the probability to find a nucleon with momentum ${\bf p}_t$ and removal
energy $E$ in the nucleus.}$^)$
of target nucleus $A$ normalized to unity;
$Z$ and $N$ are the numbers of protons and neutrons in
the target nucleus ($A=N+Z$), $M_{A}$  and $R$ are its mass and radius
\footnote{$^)$It is determined from the relation $\rho_{N}(R)=0.03\rho_{0}$ [28], and is
equal to 4.0, 5.259, 5.763, 7.128, 8.871, 9.214 fm, respectively, for $^{12}$C, $^{27}$Al, $^{40}$Ca, $^{93}$Nb, $^{208}$Pb, $^{238}$U target nuclei considered in the present work.}$^)$;
$m_N$ is the bare nucleon mass.

 Following [21--24], we assume that the off-shell cross sections
$d\sigma_{{\gamma}N\to {\eta^\prime}N}(\sqrt{s},<m^*_{{\eta^\prime}}>,{\bf p}^{\prime}_{\eta^\prime}) /d{\bf p}^{\prime}_{\eta^\prime}$ and $\sigma_{{\gamma}N\to {\eta^\prime}N}(\sqrt{s},<m^*_{{\eta^\prime}}>)$
for ${\eta^\prime}$ production in reactions (1) and (2)
are equivalent to the respective on-shell cross sections calculated for the off-shell kinematics of the elementary processes  (1), (2) as well as for the ${\eta^\prime}$ in-medium mass $<m^*_{{\eta^\prime}}>$.
Accounting for the two-body kinematics of these processes, we can
get the following expression for the former differential cross section:
\begin{equation}
\frac{d\sigma_{{\gamma}N\rightarrow {\eta^\prime}N}(\sqrt{s},<m^*_{{\eta^\prime}}>,{\bf p}^{\prime}_{\eta^\prime})}
{d{\bf p}^{\prime}_{\eta^\prime}}
={\frac{{\pi}}{I_2(s,m_N,<m^*_{{\eta^\prime}}>)E^{\prime}_{{\eta^\prime}}}}
{\frac{d\sigma_{{\gamma}N\rightarrow {\eta^\prime}N}({\sqrt{s}},<m^*_{{\eta^\prime}}>,{\theta_{\eta^\prime}^*})}
{d{\bf \Omega}_{\eta^\prime}^{*}}}\times
\end{equation}
$$
\times
{\frac{1}{(\omega+E_t)}}\delta\left[\omega+E_t-\sqrt{m_N^2+({\bf Q}+{\bf p}_t)^2}\right],
$$
where
\begin{equation}
I_2(s,m_N,<m^*_{{\eta^\prime}}>)=\frac{\pi}{2}\frac{\lambda(s,m_{N}^{2},<m^*_{{\eta^\prime}}>^{2})}{s},
\end{equation}
\begin{equation}
\lambda(x,y,z)=\sqrt{{\left[x-({\sqrt{y}}+{\sqrt{z}})^2\right]}{\left[x-
({\sqrt{y}}-{\sqrt{z}})^2\right]}},
\end{equation}
\begin{equation}
\omega=E_{\gamma}-E^{\prime}_{\eta^\prime}, \,\,\,\,{\bf Q}={\bf p}_{\gamma}-{\bf p}^{\prime}_{\eta^\prime}.
\end{equation}
Here,
$d\sigma_{{\gamma}N\rightarrow {\eta^\prime}N}({\sqrt{s}},<m^*_{{\eta^\prime}}>,{\theta_{\eta^\prime}^*})/d{\bf \Omega}_{\eta^\prime}^{*}$
is the off-shell differential cross section for the production of an
$\eta^\prime$ meson with mass $<m^*_{{\eta^\prime}}>$
under the polar angle ${\theta_{\eta^\prime}^*}$ in the ${\gamma}N$ c.m.s.

    The currently available recent experimental information on the angular distributions of the outgoing
$\eta^\prime$ mesons in the ${\gamma}p \to {\eta^\prime}p$ [25, 29--31] and ${\gamma}n \to {\eta^\prime}n$ [25]
reactions in the photon energy range $E_{\gamma} \le 2.5$ GeV can be fitted with Legendre polynomials as [25]:
\begin{equation}
\frac{d\sigma_{{\gamma}p\rightarrow {\eta^\prime}p}({\sqrt{s}},m_{{\eta^\prime}},{\theta_{\eta^\prime}^*})}{d{\bf \Omega}_{\eta^\prime}^{*}}
=\left\{
\begin{array}{ll}
	\frac{\sigma_{{\gamma}p\to {\eta^\prime}p}(\sqrt{s},m_{{\eta^\prime}})}{4\pi}
	&\mbox{for $E^{\rm thr}_{\gamma}(m_{{\eta^\prime}}) \le E_{\gamma}< 1.525~{\rm GeV}$}, \\
	&\\
 \frac{p^*_{\eta^\prime}(m_{\eta^\prime})}{p^*_{\gamma}(m^2_N)}\sum_{i=0}^3 A_i^{{\gamma}p\rightarrow {\eta^\prime}p}P_i(\cos{{\theta_{\eta^\prime}^*}})
	&\mbox{for $1.525~{\rm GeV} \le E_{\gamma}\le 2.5~{\rm GeV}$},
\end{array}
\right.	
\end{equation}
where
\begin{equation}
\sigma_{{\gamma}p\to {\eta^\prime}p}(\sqrt{s},m_{{\eta^\prime}})=\left\{
\begin{array}{ll}
	3.856\left[(E_{\gamma}-E_{\gamma}^{\rm thr}(m_{{\eta^\prime}}))/{\rm GeV}\right]^{0.5}~[{\rm {\mu}b}]
	&\mbox{for $E^{\rm thr}_{\gamma}(m_{{\eta^\prime}}) \le E_{\gamma}< 1.51~{\rm GeV}$}, \\
	&\\
                   26.602-16.973(E_{\gamma}/{\rm GeV})~[{\rm {\mu}b}]
	&\mbox{for $1.51~{\rm GeV} \le E_{\gamma} < 1.525~{\rm GeV}$},
\end{array}
\right.	
\end{equation}
\begin{equation}
A_0^{{\gamma}p\rightarrow{\eta^\prime}p}=0.751(E_{\gamma}/{\rm GeV})^{-2.908}~[{\rm {\mu}b}/{\rm sr}],
\end{equation}
\begin{equation}
A_1^{{\gamma}p\rightarrow{\eta^\prime}p}=\left\{
\begin{array}{ll}
	0.0021(E_{\gamma}/{\rm GeV})^{6.305}~[{\rm {\mu}b}/{\rm sr}]
	&\mbox{for $1.525~{\rm GeV} \le E_{\gamma}< 1.9~{\rm GeV}$}, \\
	&\\
                   1.3305(E_{\gamma}/{\rm GeV})^{-3.748}~[{\rm {\mu}b/{\rm sr}}]
	&\mbox{for $1.9~{\rm GeV} \le E_{\gamma} \le 2.5~{\rm GeV}$},
\end{array}
\right.	
\end{equation}
\begin{equation}
A_2^{{\gamma}p\rightarrow{\eta^\prime}p}=\left\{
\begin{array}{lll}
	-0.02~[{\rm {\mu}b}/{\rm sr}]
	&\mbox{for $1.525~{\rm GeV} \le E_{\gamma}< 1.7~{\rm GeV}$}, \\
	&\\
        -0.53+0.3(E_{\gamma}/{\rm GeV})~[{\rm {\mu}b/{\rm sr}}]
	&\mbox{for $1.7~{\rm GeV} \le E_{\gamma} < 2.1~{\rm GeV}$}, \\
    &\\
    0.88(E_{\gamma}/{\rm GeV})^{-2.93}~[{\rm {\mu}b/{\rm sr}}]
    &\mbox{for $2.1~{\rm GeV} \le E_{\gamma} \le 2.5~{\rm GeV}$},
\end{array}
\right.	
\end{equation}
\begin{equation}
A_3^{{\gamma}p\rightarrow{\eta^\prime}p}=\left\{
\begin{array}{ll}
	0.1\left[(E_{\gamma}/{\rm GeV})-2.0\right]~[{\rm {\mu}b}/{\rm sr}]
	&\mbox{for $1.525~{\rm GeV} \le E_{\gamma}< 2.0~{\rm GeV}$}, \\
	&\\
                   0
	&\mbox{for $2.0~{\rm GeV} \le E_{\gamma} \le 2.5~{\rm GeV}$}
\end{array}
\right.	
\end{equation}
and
\begin{equation}
\frac{d\sigma_{{\gamma}n\rightarrow {\eta^\prime}n}({\sqrt{s}},m_{{\eta^\prime}},{\theta_{\eta^\prime}^*})}{d{\bf \Omega}_{\eta^\prime}^{*}}
=\left\{
\begin{array}{ll}
	\frac{d\sigma_{{\gamma}p\rightarrow {\eta^\prime}p}({\sqrt{s}},m_{{\eta^\prime}},{\theta_{\eta^\prime}^*})}{d{\bf \Omega}_{\eta^\prime}^{*}}
	&\mbox{for $E^{\rm thr}_{\gamma}(m_{{\eta^\prime}}) \le E_{\gamma}< 1.525~{\rm GeV}$}, \\
	&\\
 \frac{p^*_{\eta^\prime}(m_{\eta^\prime})}{p^*_{\gamma}(m^2_N)}\sum_{i=0}^3 A_i^{{\gamma}n\rightarrow {\eta^\prime}n}P_i(\cos{{\theta_{\eta^\prime}^*}})
	&\mbox{for $1.525~{\rm GeV} \le E_{\gamma}\le 2.5~{\rm GeV}$},
\end{array}
\right.	
\end{equation}
\begin{equation}
A_0^{{\gamma}n\rightarrow{\eta^\prime}n}=\left\{
\begin{array}{lll}
	4.278(E_{\gamma}/{\rm GeV})^{-7.277}~[{\rm {\mu}b/{\rm sr}}]
	&\mbox{for $1.525~{\rm GeV} \le E_{\gamma}< 1.7~{\rm GeV}$}, \\
	&\\
        0.204(E_{\gamma}/{\rm GeV})^{-1.546}~[{\rm {\mu}b/{\rm sr}}]
	&\mbox{for $1.7~{\rm GeV} \le E_{\gamma} < 2.2~{\rm GeV}$}, \\
    &\\
    1.287(E_{\gamma}/{\rm GeV})^{-3.888}~[{\rm {\mu}b/{\rm sr}}]
    &\mbox{for $2.2~{\rm GeV} \le E_{\gamma} \le 2.5~{\rm GeV}$},
\end{array}
\right.	
\end{equation}
\begin{equation}
A_i^{{\gamma}n\rightarrow{\eta^\prime}n}=A_i^{{\gamma}p\rightarrow{\eta^\prime}p},\,\,\, i=1,2,3.
\end{equation}
Here,
\begin{equation}
E^{\rm thr}_{\gamma}(m_{{\eta^\prime}})=\frac{m_{{\eta^\prime}}(2m_N+m_{{\eta^\prime}})}{2m_N},
\end{equation}
\begin{equation}
p^*_{\eta^\prime}(m_{\eta^\prime})=\frac{1}{2\sqrt{s}}\lambda(s,m_{N}^2,m_{\eta^\prime}^{2}),
\end{equation}
\begin{equation}
p_{\gamma}^*(m_{N}^2)=\frac{1}{2\sqrt{s}}\lambda(s,0,m_{N}^2).
\end{equation}
When the reaction ${\gamma}N\to {\eta^\prime}N$ occurs on an
off-shell target nucleon, then instead of the incident photon energy $E_{\gamma}$, appearing in the Eqs. (20)--(27),
and of the nucleon mass squared $m_N^2$ in the Eq. (31), we should use, respectively,
the effective energy $E_{\gamma}^{\rm eff}$ and the quantity $E_{t}^{2}-p_{t}^2$.
The latter one should be calculated in line with Eq. (15). The energy $E_{\gamma}^{\rm eff}$
can be expressed in terms of the collision energy squared $s$, defined by Eq. (14), as follows:
\begin{equation}
E_{\gamma}^{\rm eff}=\frac{s-m_N^2}{2m_N}.
\end{equation}
The $\eta^\prime$ meson production angle ${\theta_{\eta^\prime}^*}$ in the ${\gamma}N$ c.m.s.
in this case and in the $\eta^\prime$ medium modification scenario
is defined by:
\begin{equation}
\cos{\theta_{\eta^\prime}^*}=\frac{{\bf p}_{\gamma}^*(E_{t}^{2}-p_{t}^2){\bf p}_{\eta^\prime}^*(<m^*_{{\eta^\prime}}>)}{p_{\gamma}^*(E_{t}^{2}-p_{t}^2)p_{\eta^\prime}^*(<m^*_{{\eta^\prime}}>)}.
\end{equation}
Writing the invariant $t$--the squared four-momentum transfer between the incident photon and the secondary
$\eta^\prime$ meson--in the l.s. and in the ${\gamma}N$ c.m.s. and equating the results, we obtain:
\begin{equation}
\cos{\theta_{\eta^\prime}^*}=\frac{p_{\gamma}p^{\prime}_{\eta^\prime}\cos{\theta^{\prime}_{\eta^\prime}}+
(E_{\gamma}^*E_{\eta^\prime}^*-E_{\gamma}E^{\prime}_{\eta^\prime})}
{p_{\gamma}^*(E_{t}^{2}-p_{t}^2)p_{\eta^\prime}^*(<m^*_{\eta^\prime}>)},
\end{equation}
where
\begin{equation}
E_{\gamma}^*=p_{\gamma}^*(E_{t}^{2}-p_{t}^2),\,\,\,\,
E_{\eta^\prime}^*=\sqrt{[p_{\eta^\prime}^*(<m^*_{{\eta^\prime}}>)]^2+(<m^*_{{\eta^\prime}}>)^2}.
\end{equation}
In Eq. (34), $\theta^{\prime}_{\eta^\prime}$
is the angle between the initial photon three-momentum ${\bf p}_{\gamma}$ and the in-medium $\eta^\prime$
three-momentum ${\bf p}^{\prime}_{\eta^\prime}$ in the l.s. frame. For the sake of numerical simplicity,
we will assume that the direction of the $\eta^\prime$ meson three-momentum remains unchanged as it propagates
from its creation point inside the nucleus to the vacuum far away from the nucleus. As a consequence, this angle
does not deviate from the angle  $\theta_{\eta^\prime}$ between the incident photon momentum ${\bf p}_{\gamma}$
and vacuum $\eta^\prime$ momentum ${\bf p}_{\eta^\prime}$ in the l.s. In this case, the quantity
$d{\bf p}^{\prime}_{\eta^\prime}/d{\bf p}_{\eta^\prime}$, entering into Eq. (9), can be put in the
simple form $p^{\prime}_{\eta^\prime}/p_{\eta^\prime}$.

       For the $\eta^\prime$ production calculations in the case of  $^{12}$C and $^{27}$Al, $^{40}$Ca, $^{93}$Nb, $^{208}$Pb, $^{238}$U target nuclei reported here
we have employed for the nuclear density $\rho({\bf r})$,
respectively, the harmonic oscillator and the Woods-Saxon distributions, given in [32].
For the latter one, we take
$R_{1/2}=3.347~{\rm fm}$ for $^{27}$Al, $R_{1/2}=3.852~{\rm fm}$ for $^{40}$Ca,
$R_{1/2}=5.216~{\rm fm}$ for $^{93}$Nb, $R_{1/2}=6.959~{\rm fm}$ for $^{208}$Pb, $R_{1/2}=7.302~{\rm fm}$ for $^{238}$U.
The nuclear spectral function $P_A({\bf p}_t,E)$ for the $^{12}$C target nucleus was taken
 from [21]. The single-particle part of this function for $^{27}$Al, $^{40}$Ca, $^{93}$Nb and $^{238}$U
target nuclei was assumed to be the same as that for $^{208}$Pb.
The latter was taken from [22]. The correlated part of the
nuclear spectral function for these target nuclei was borrowed from [21].

     The absorption cross section $\sigma_{{\eta^\prime}N}$ can be extracted, for example, from
a comparison of the calculations with the transparency ratio of the $\eta^\prime$ meson, normalized to carbon,
as measured in [19]
\begin{equation}
T_A=\frac{12~\sigma_{{\gamma}A \to {\eta^\prime}X}}{A~\sigma_{{\gamma}C \to {\eta^\prime}X}}.
\end{equation}
Here, $\sigma_{{\gamma}A \to {\eta^\prime}X}$ and
$\sigma_{{\gamma}C \to {\eta^\prime}X}$ are inclusive total cross sections for $\eta^\prime$ production in
${\gamma}A$ and ${\gamma}{\rm C}$ collisions, respectively. If the primary photon--induced reaction channels
(1), (2) dominate in the $\eta^\prime$ production in ${\gamma}A$ interactions, then, according to (10)
and (11), we have:
\begin{equation}
T_A=\frac{12~I_V[A]}{A~I_V[C]}
\frac{\left[\frac{Z}{A}\left<\sigma_{{\gamma}p\to {\eta^\prime}p}({\bf p}_{\gamma})\right>_A+
\frac{N}{A}\left<\sigma_{{\gamma}n\to {\eta^\prime}n}({\bf p}_{\gamma})\right>_A\right]}
{\left[\frac{1}{2}\left<\sigma_{{\gamma}p\to {\eta^\prime}p}({\bf p}_{\gamma})\right>_C+
\frac{1}{2}\left<\sigma_{{\gamma}n\to {\eta^\prime}n}({\bf p}_{\gamma})\right>_C\right]}.
\end{equation}
Ignoring the medium and isospin effects
\footnote{$^)$It is worth noting that these effects lead to only small corrections to the ratio
$T_A$. They are within several percent, as our calculations by (37), (38) for the nucleus with a
diffuse boundary showed.}$^)$,
from (37) we approximately obtain:
\begin{equation}
T_A \approx \frac{12~I_V[A]}{A~I_V[C]}.
\end{equation}
The integral (11) for $I_V[A]$ in the case of a nucleus of a radius $R=r_0A^{1/3}$ with a sharp
boundary has the following simple form:
\begin{equation}
I_V[A]=\frac{3A}{2a_1}\left\{1-\frac{2}{a_1^2}[1-(1+a_1)e^{-a_1}]\right\},
\end{equation}
where $a_1=2R/{\lambda}_{\eta^\prime}$ and ${\lambda}_{\eta^\prime}=1/({\rho_0}\sigma_{{\eta^\prime}N})$.
In particular, it can be easily obtained from the more general expression (28) given in [33] in the limit
$a_2 \to 0$. The transparency ratio ${\tilde T_A}$ can be defined also as the ratio of the inclusive nuclear
$\eta^\prime$ photoproduction cross section divided by $A$ times the same quantity on a free nucleon [34].
According to (10) and (39), for this quantity one gets:
\begin{equation}
{\tilde T_A}=\frac{1}{A}I_V[A]=\frac{{\pi}R^2}{A\sigma_{{\eta^\prime}N}}
\left\{1+\left(\frac{\lambda_{\eta^\prime}}{R}\right)e^{-\frac{2R}{\lambda_{\eta^\prime}}}+
\frac{1}{2}\left(\frac{\lambda_{\eta^\prime}}{R}\right)^2\left[e^{-\frac{2R}{\lambda_{\eta^\prime}}}-1\right]\right\}.
\end{equation}
This formula was derived also in [34] and was reported in [19] in ${\eta^\prime}$ photoproduction
from nuclei.

\section*{2.2. Two-step ${\eta^\prime}$ production mechanism}

\hspace{1.5cm} Kinematical considerations show that in the incident photon energy range
of our interest the following two-step production process
may contribute to the ${\eta^\prime}$ creation in ${\gamma}A$ interactions. An incident photon
can produce in the first inelastic collision with an intranuclear nucleon also a pion through the
elementary reaction [24]
\begin{equation}
\gamma+N_1 \to 2\pi+N.
\end{equation}
Then, the intermediate pion, which is assumed to be on-shell, produces the ${\eta^\prime}$ meson
on another nucleon of the target nucleus via the elementary subprocess with the lowest free
production threshold momentum (1.43 GeV/c):
\begin{equation}
\pi+N_2 \to {\eta^\prime}+N.
\end{equation}
It should be noted that the elementary processes ${\pi}N \to {\eta^\prime}N{\pi}$ with one
pion in final states is expected to play a minor role in ${\eta^\prime}$ production in
${\gamma}A$ reactions for kinematics of interest. This is due to the following. For instance,
at beam energy of 2 GeV the maximum allowable momentum of a pion in the process (41) taking place
on a free nucleon at rest is 1.85 GeV/c [24]. This momentum is close to the threshold momentum of
1.72 GeV/c of the channel ${\pi}N \to {\eta^\prime}N{\pi}$. Therefore, it is energetically suppressed.

     Taking into account the medium effects on the ${\eta^\prime}$ mass on the same footing as that
employed in calculating the ${\eta^\prime}$ production cross sections (9), (10) from the primary
processes (1), (2) and ignoring the distortion of the initial photon inside the target nucleus as well as
using the results given in [23, 24], we get the following expression for the
${\eta^\prime}$ production total cross section for ${\gamma}A$ reactions
from the secondary channel (42):
\begin{equation}
\sigma_{{\gamma}A\to {\eta^\prime}X}^{({\rm sec})}
(E_{\gamma})=I_{V}^{({\rm sec})}[A]
\sum_{\pi'=\pi^+,\pi^0,\pi^-}\int\limits_{4\pi}d{\bf \Omega}_{\pi}
\int\limits_{p_{\pi}^{{\rm abs}}}^{p_{\pi}^{{\rm lim}}
(\vartheta_{\pi})}p_{\pi}^{2}dp_{\pi}
\times
\end{equation}
$$
\times
\left[\frac{Z}{A}\left<\frac{d\sigma_{{\gamma}p\to {\pi'}X}({\bf p}_{\gamma},
{\bf p}_{\pi})}{d{\bf p}_{\pi}}\right>_A+
\frac{N}{A}\left<\frac{d\sigma_{{\gamma}n\to {\pi'}X}({\bf p}_{\gamma},
{\bf p}_{\pi})}{d{\bf p}_{\pi}}\right>_A\right]
\times
$$
$$
\times
\left[\frac{Z}{A}\left<\sigma_{{\pi'}p\to {\eta^\prime}N}({\bf p}_{\pi})\right>_A+
\frac{N}{A}\left<\sigma_{{\pi'}n\to {\eta^\prime}N}({\bf p}_{\pi})\right>_A\right],
$$
where
\footnote{$^)$It should be pointed out that due to the very moderate dependence of the quantity
$I_{V}^{({\rm sec})}[A]$ on the $\eta^\prime$ production angle in the lab frame, we assume writing
the formula (44) for it that the $\eta^\prime$ meson moves in the nucleus substantially in forward
direction.}$^)$
\begin{equation}
I_{V}^{({\rm sec})}[A]=2{\pi}A^2\int\limits_{0}^{R}r_{\bot}dr_{\bot}
\int\limits_{-\sqrt{R^2-r_{\bot}^2}}^{\sqrt{R^2-r_{\bot}^2}}dz
\rho(\sqrt{r_{\bot}^2+z^2})
\int\limits_{0}^{\sqrt{R^2-r_{\bot}^2}-z}dl
\rho(\sqrt{r_{\bot}^2+(z+l)^2})
\times
\end{equation}
$$
\times
\exp{\left[-\sigma_{{\pi}N}^{{\rm tot}}A\int\limits_{z}^{z+l}
\rho(\sqrt{r_{\bot}^2+x^2})dx
-\sigma_{{\eta^\prime}N}A\int\limits_{z+l}^{\sqrt{R^2-r_{\bot}^2}}
\rho(\sqrt{r_{\bot}^2+x^2})dx\right]},
$$
\begin{equation}
\left<\frac{d\sigma_{{\gamma}N\to {\pi'}X}({\bf p}_{\gamma},
{\bf p}_{\pi})}
{d{\bf p}_{\pi}}\right>_A=
\int\int
P_A({\bf p}_t,E)d{\bf p}_tdE
\left[\frac{d\sigma_{{\gamma}N\to {\pi'}X}(\sqrt{s},{\bf p}_{\pi})}
{d{\bf p}_{\pi}}\right],
\end{equation}
\begin{equation}
\left<\sigma_{{\pi'}N\to {\eta^\prime}N}({\bf p}_{\pi})\right>_A=
\int\int
P_A({\bf p}_t,E)d{\bf p}_tdE
\left[\sigma_{{\pi'}N\to {\eta^\prime}N}(\sqrt{s_1},<m^*_{{\eta^\prime}}>)\right];
\end{equation}
\begin{equation}
  s_1=(E_{\pi}+E_{t})^2-({\bf p}_{\pi}+{\bf p}_{t})^2,
\end{equation}
\begin{equation}
 p_{\pi}^{{\rm lim}}(\vartheta_{\pi}) =
\frac{{\beta}_{A}p_{\gamma}\cos{\vartheta_{\pi}}+
 (E_{\gamma}+M_A)\sqrt{{\beta}_{A}^2-4m_{\pi}^{2}(s_{A}+
p_{\gamma}^{2}\sin^{2}{\vartheta_{\pi}})}}{2(s_{A}+
p_{\gamma}^{2}\sin^{2}{\vartheta_{\pi}})},
\end{equation}
\begin{equation}
 {\beta}_A=s_{A}+m_{\pi}^{2}-M_{A}^{2},\,\,s_A=(E_{\gamma}+M_A)^2-p_{\gamma}^{2},
\end{equation}
\begin{equation}
\cos{\vartheta_{\pi}}={\bf \Omega}_{\gamma}{\bf \Omega}_{\pi},\,\,\,\,
{\bf \Omega}_{\gamma}={\bf p}_{\gamma}/p_{\gamma},\,\,\,\,{\bf \Omega}_{\pi}={\bf p}_{\pi}/p_{\pi}.
\end{equation}
Here, $d\sigma_{{\gamma}N\to {\pi'}X}(\sqrt{s},{\bf p}_{\pi})/d{\bf p}_{\pi}$ are the
inclusive differential cross sections for pion production from
the primary photon-induced reaction channel (41);
$\sigma_{{\pi'}N\to {\eta^\prime}N}(\sqrt{s_1},<m^*_{{\eta^\prime}}>)$ are the in-medium total
cross sections for the production of $\eta^\prime$ mesons with the reduced mass $<m^*_{{\eta^\prime}}>$
in ${\pi}^{\prime}N$ collisions via the subprocess (42) at the ${\pi}^{\prime}N$ center-of-mass energy
$\sqrt{s_1}$; $\sigma_{{\pi}N}^{{\rm tot}}$ is the total cross section of the free ${\pi}N$
interaction; ${\bf p}_{\pi}$ and $E_{\pi}$ are the three-momentum and total energy
($E_{\pi}=\sqrt{{\bf p}_{\pi}^2+m_{\pi}^2}$) of an intermediate pion; $p_{\pi}^{\rm abs}$ is the
absolute threshold momentum for the $\eta^\prime$ production on the residual nucleus by this pion.

   We will calculate hereafter the differential cross sections
$d\sigma_{{\gamma}N\to {\pi'}X}(\sqrt{s},{\bf p}_{\pi})/d{\bf p}_{\pi}$, entering into Eqs. (43), (45),
following strictly the approach [24]. The elementary $\eta^\prime$ production reactions
${\pi}^+n \to {\eta^\prime}p$,
${\pi}^0p \to {\eta^\prime}p$, ${\pi}^0n \to {\eta^\prime}n$ and
${\pi}^-p \to {\eta^\prime}n$ have been included in our calculations
of the $\eta^\prime$ production on nuclei.
Isospin considerations show that the following relations among the total cross sections of
these reactions exist:
\begin{equation}
\sigma_{{\pi}^{+}n \to {\eta^\prime}p}=\sigma_{{\pi}^{-}p \to {\eta^\prime}n},
\end{equation}
\begin{equation}
\sigma_{{\pi}^{0}p \to {\eta^\prime}p}=\sigma_{{\pi}^{0}n \to {\eta^\prime}n}=\frac{1}{2}\sigma_{{\pi}^{-}p \to {\eta^\prime}n}.
\end{equation}
For the free total cross section
$\sigma_{{\pi}^{-}p \to {\eta^\prime}n}$ we have used the following parametrization:
\begin{equation}
\sigma_{{\pi}^-p \to {\eta^\prime}n}({\sqrt{s_1}},m_{\eta^\prime})=\left\{
\begin{array}{ll}
	223.5\left(\sqrt{s_1}-\sqrt{s_0}\right)^{0.352}~[{\rm {\mu}b}]
	&\mbox{for $0<\sqrt{s_1}-\sqrt{s_0}\le 0.354~{\rm GeV}$}, \\
	&\\
                   35.6/\left(\sqrt{s_1}-\sqrt{s_0}\right)^{1.416}~[{\rm {\mu}b}]
	&\mbox{for $\sqrt{s_1}-\sqrt{s_0} > 0.354~{\rm GeV}$},
\end{array}
\right.	
\end{equation}
where $\sqrt{s_0}=m_{\eta^\prime}+m_N$ is the threshold energy.
The comparison of the results of calculations by (53) (solid line) with the available experimental
data [35] for the ${\pi}^{-}p \to {\eta^\prime}n$ (full triangles) and for the
${\pi}^{+}n \to {\eta^\prime}p$ (full circles) reactions is shown in figure 1. It is seen that the
parametrization (53) fits well the existing set of data for these reactions.
\begin{figure}[htb]
\begin{center}
\includegraphics[width=12.0cm]{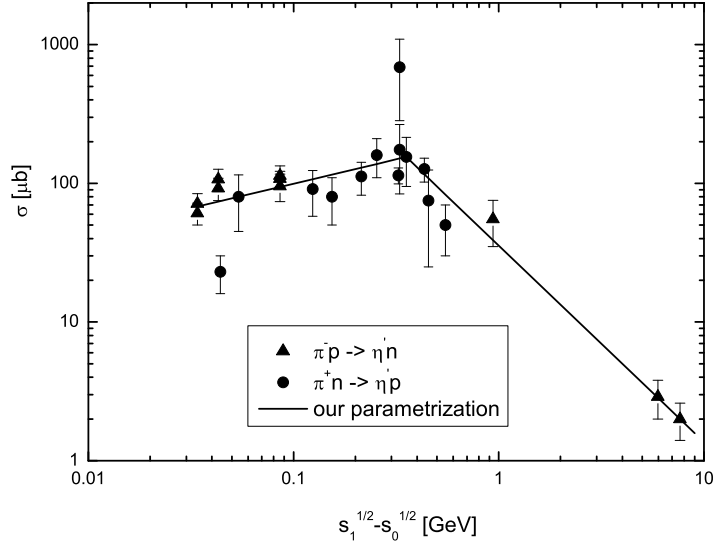}
\vspace*{-2mm} \caption{Total cross sections for the $\eta^\prime$ production in the
${\pi}^{-}p \to {\eta^\prime}n$ and ${\pi}^{+}n \to {\eta^\prime}p$ reactions. For notation see the text.}
\label{void}
\end{center}
\end{figure}

  Now, we discuss the results of our calculations for ${\eta^\prime}$
  production in ${\gamma}A$ interactions within the model outlined above.

\section*{3. Results and discussion}

\hspace{1.5cm} First of all, we consider the A--dependences of the total $\eta^\prime$ production
cross sections from the one-step and two-step $\eta^\prime$ creation mechanisms
in ${\gamma}A$ ($A=$$^{12}$C, $^{27}$Al, $^{40}$Ca, $^{93}$Nb, $^{208}$Pb, and $^{238}$U) collisions calculated
for $E_{\gamma}=1.85$ GeV on the basis of Eqs. (10) and (43). These dependences are shown in figure 2.
\begin{figure}[htb]
\begin{center}
\includegraphics[width=12.0cm]{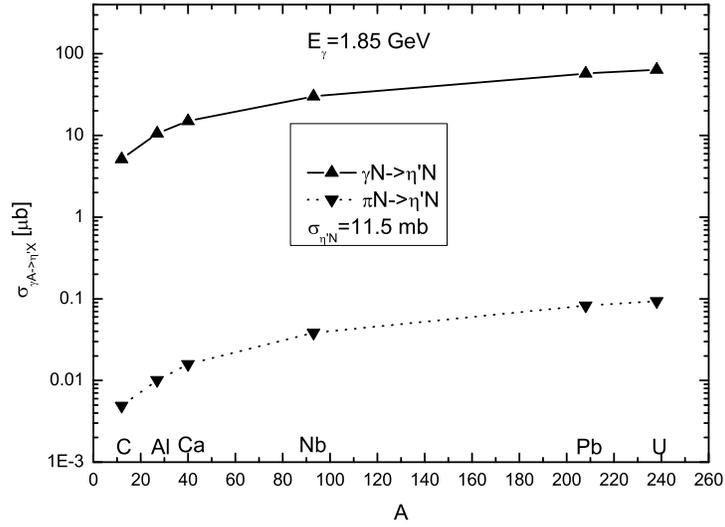}
\vspace*{-2mm} \caption{A--dependences of the total cross sections of $\eta^\prime$ production
by 1.85 GeV photons from primary ${\gamma}N \to {\eta^\prime}N$ channels (1), (2) and from secondary
${\pi}N \to {\eta^\prime}N$ processes (42) in the full phase space in the scenario without $\eta^\prime$
mass shift. The absorption of $\eta^\prime$ mesons in nuclear matter was determined assuming an inelastic cross
section $\sigma_{{\eta^\prime}N}=11.5$ mb. The lines are included to guide the eyes.}
\label{void}
\end{center}
\end{figure}
It can be seen that the role of the secondary pion--induced reaction channel
${\pi}N \to {\eta^\prime}N$ is negligible compared to that of the primary ${\gamma}N \to {\eta^\prime}N$
processes for all considered target nuclei, which is in line with the experimental findings of [19].
This gives confidence to us that the channel ${\pi}N \to {\eta^\prime}N$ can be ignored in our
subsequent calculations.
\begin{figure}[!h]
\begin{center}
\includegraphics[width=12.0cm]{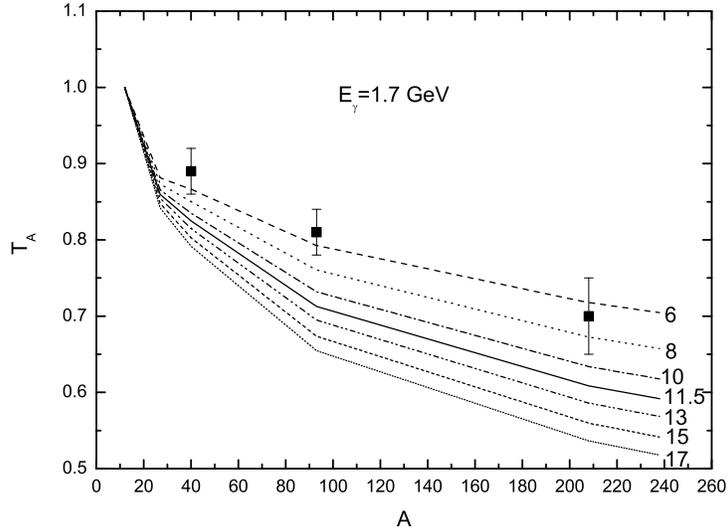}
\vspace*{-2mm} \caption{Transparency ratio $T_A$ for $\eta^\prime$ mesons as a function of the
nuclear mass number $A$ for their different in-medium absorption cross sections (in mb) at an incident
photon energy of 1.7 GeV. For notation see the text.}
\label{void}
\end{center}
\end{figure}
\begin{figure}[!h]
\begin{center}
\includegraphics[width=12.0cm]{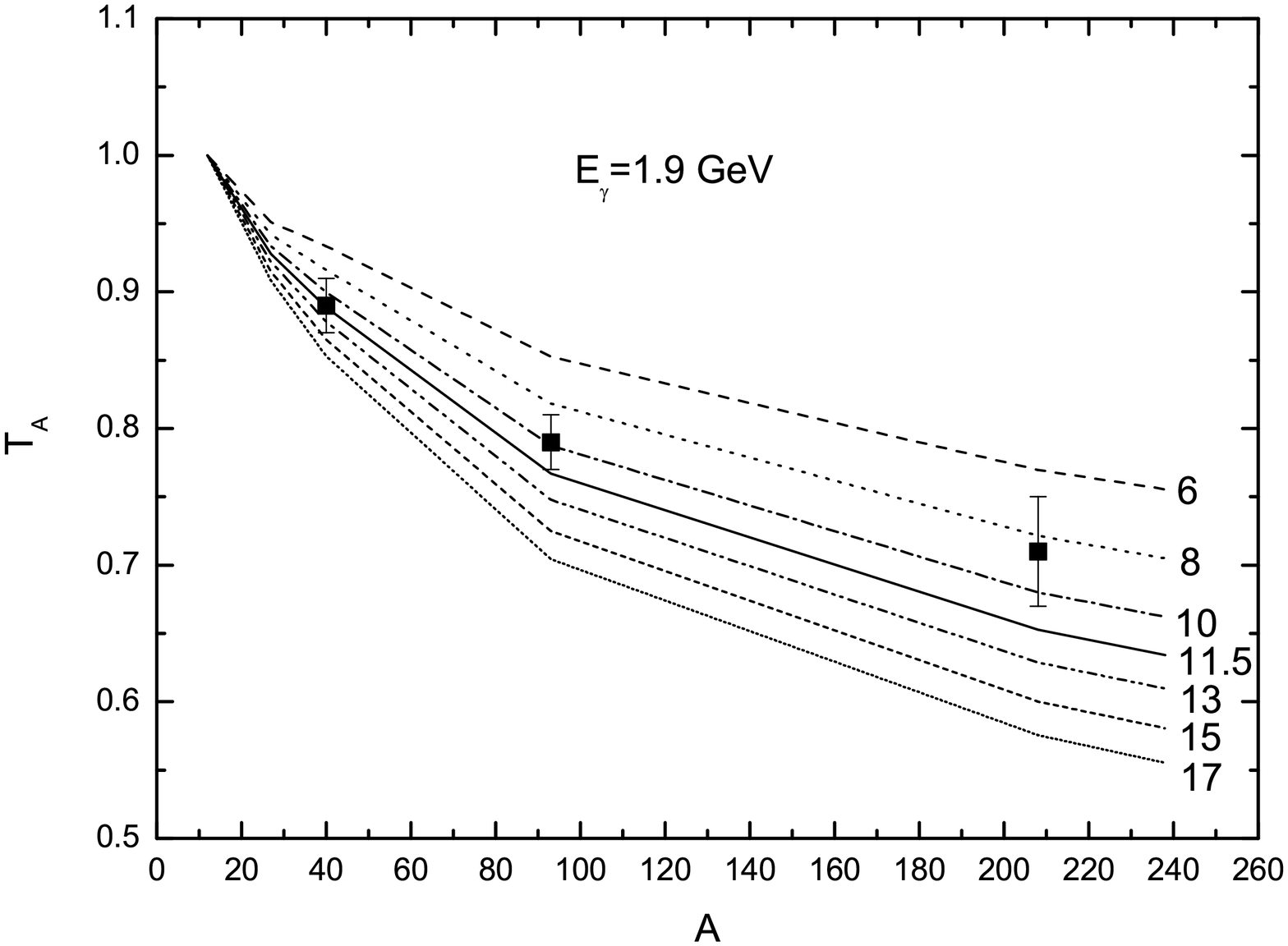}
\vspace*{-2mm} \caption{The same as in figure 3, but for the interaction of
1.9 GeV photons with the considered target nuclei.}
\label{void}
\end{center}
\end{figure}
\begin{figure}[!h]
\begin{center}
\includegraphics[width=12.0cm]{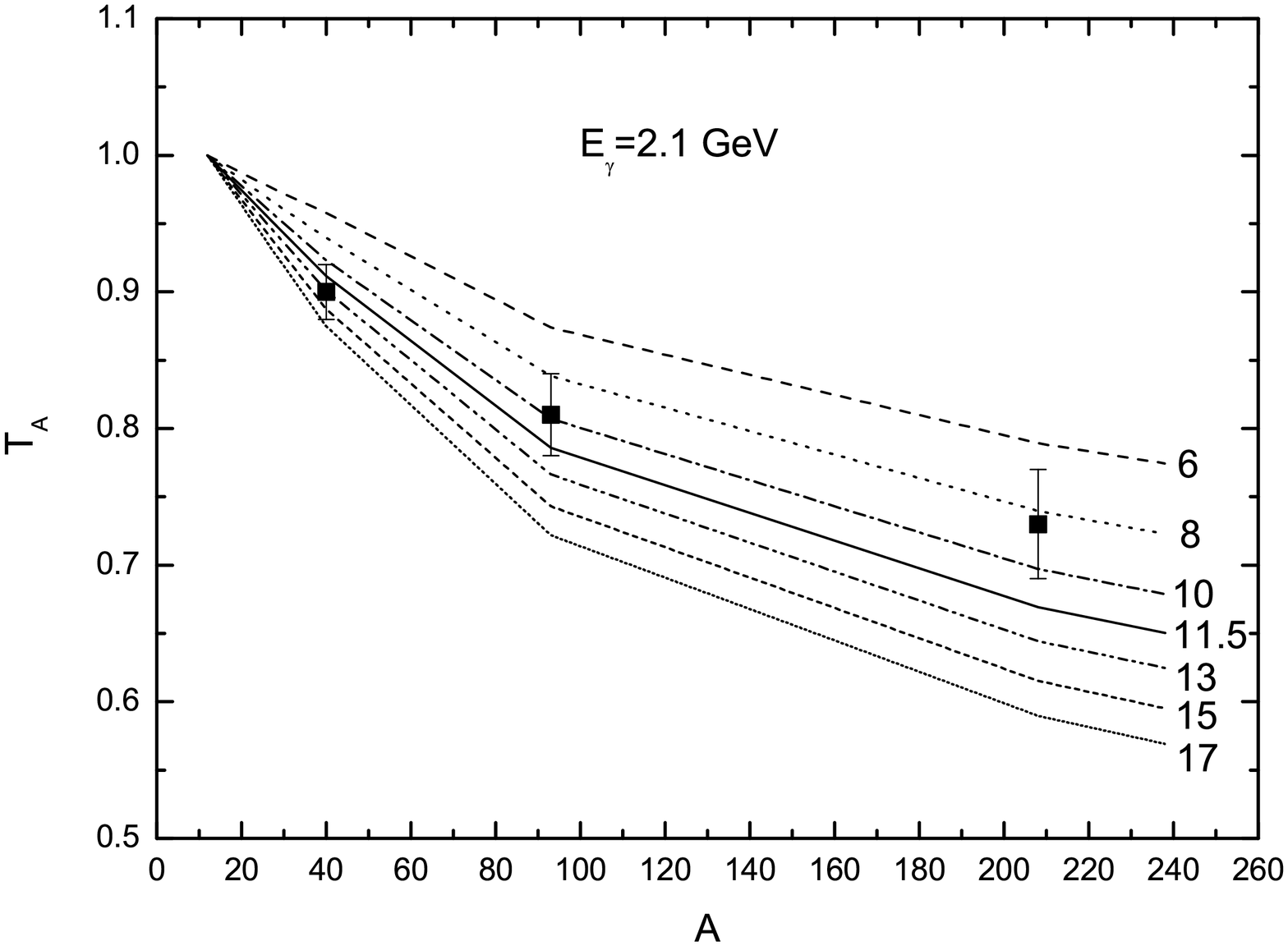}
\vspace*{-2mm} \caption{The same as in figure 3, but for the interaction of
2.1 GeV photons with the considered target nuclei.}
\label{void}
\end{center}
\end{figure}
\begin{figure}[!h]
\begin{center}
\includegraphics[width=12.0cm]{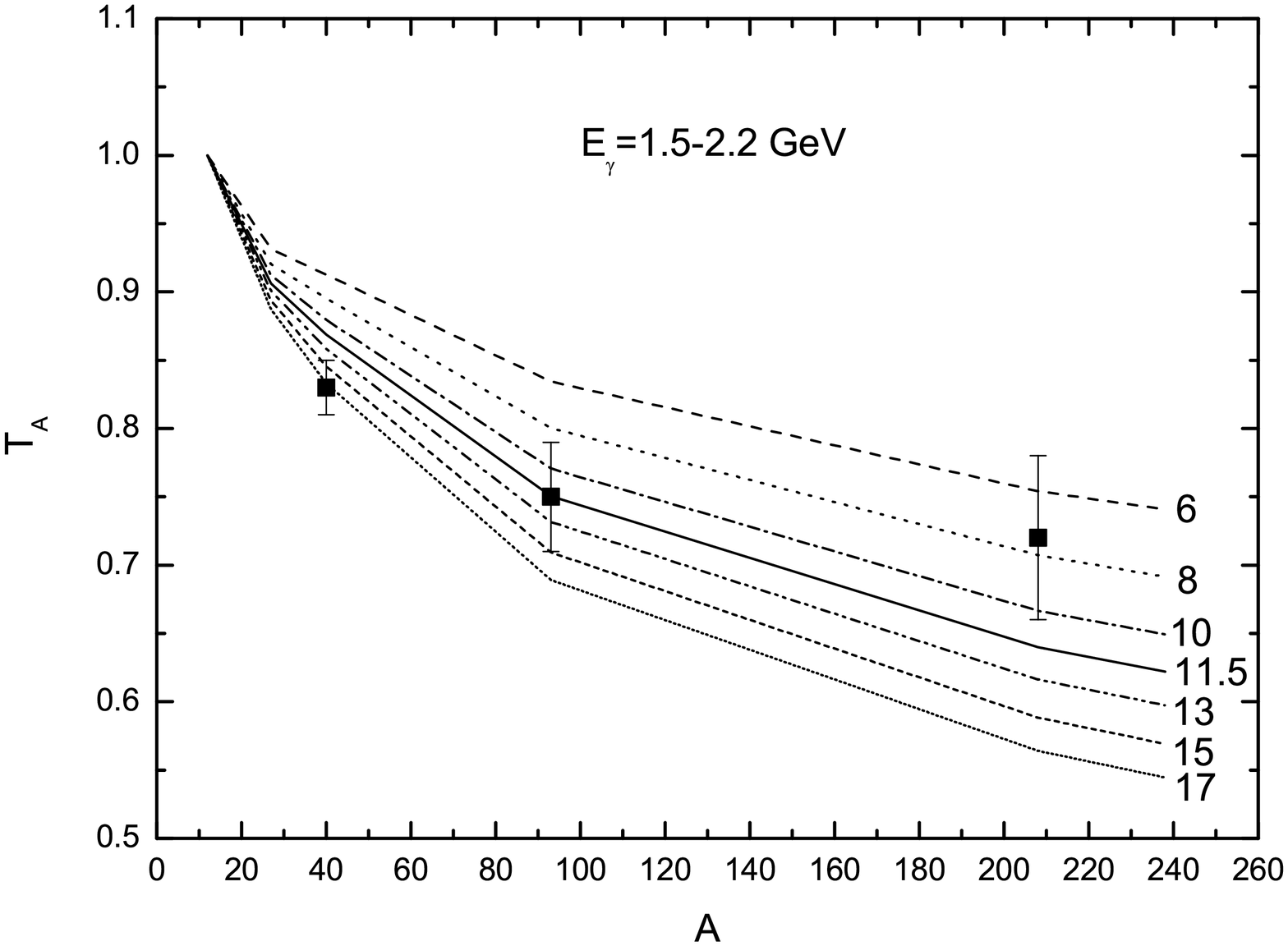}
\vspace*{-2mm} \caption{The same as in figure 3, but for the interaction of
photons of energies of 1.5--2.2 GeV with the considered target nuclei.}
\label{void}
\end{center}
\end{figure}

     Now, we focus upon the relative observable--the transparency ratio $T_A$ defined by Eq. (36).
This ratio as a function of the nuclear mass number $A$ is shown in figures 3, 4, 5 and 6 for three
initial photon energies of 1.7, 1.9, 2.1 GeV and for the photon energy range of 1.5--2.2 GeV, respectively.
The experimental data in these figures are from [19]. The curves are calculations by Eq. (37)
\footnote{$^)$It should be noted that in the case, when the incident photon energy $E_{\gamma}$
belongs to the range of 1.5--2.2 GeV, the cross sections
$\left<\sigma_{{\gamma}N\to {\eta^\prime}N}({\bf p}_{\gamma})\right>_{A,{\rm C}}$,
entering into Eq. (37), were averaged over this energy as follows:
$\int\limits_{1.5~{\rm GeV}}^{2.2~{\rm GeV}} dE_{\gamma}
\left<\sigma_{{\gamma}N\to {\eta^\prime}N}({\bf p}_{\gamma})\right>_{A,{\rm C}}/
(2.2~{\rm GeV}-1.5~{\rm GeV})$. Evidently, this treatment is absolutely correct for the description
both the data on the $\eta^\prime$ transparency ratio, given in figures 6, 9--11, and the data on the
$\eta^\prime$ momentum distribution on the $^{12}$C target, presented in figure 8. But for the
comparison to the absolute experimental cross sections, determined for this energy range,
it is needed , strictly speaking, to average the calculated cross sections over the photon energy
$E_{\gamma}$ with $1/E_{\gamma}$ weighting.}$^)$
for different values (in mb) of the in-medium inelastic cross section $\sigma_{{\eta^\prime}N}$,
as indicated by the curves, and for no $\eta^\prime$ mass shift
\footnote{$^)$It should be pointed out that the inclusion of the considered $\eta^\prime$ mass
shift (6) leads, as our calculations showed, to a change of the ratio $T_A$ only by about several
percent (see also figures 9--11 below). This gives confidence to us that the inelastic
${\eta^\prime}N$ cross section estimated from the comparison of the measured and calculated
transparency ratios is sufficiently reliable.}$^)$.
A comparison of the data to our model calculations yields $\sigma_{{\eta^\prime}N} \approx$ 6--10 mb.
This value coincides with that deduced in [19] in a low-density approximation.
\begin{figure}[!h]
\begin{center}
\includegraphics[width=12.0cm]{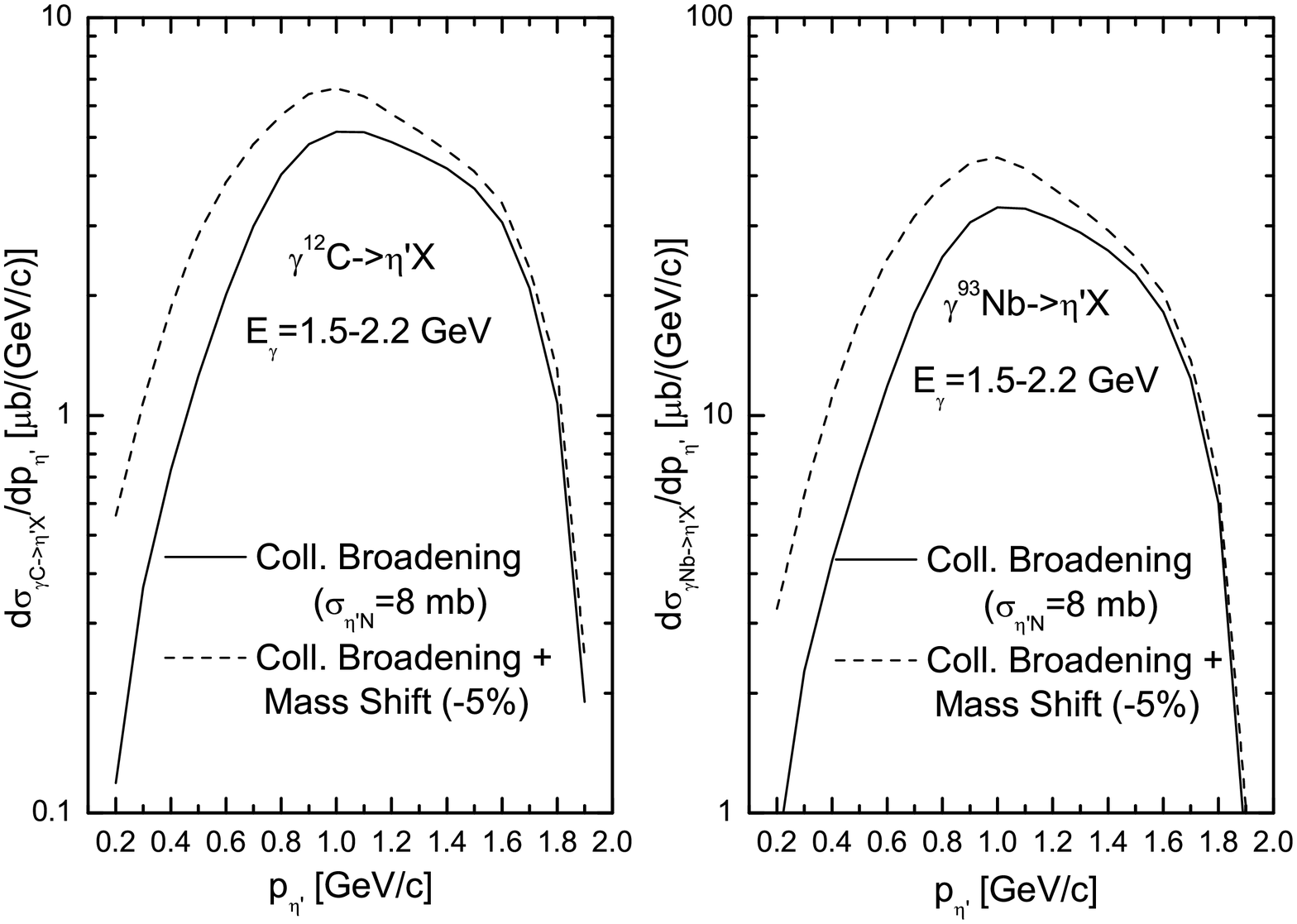}
\vspace*{-2mm} \caption{Momentum differential cross sections for the production of $\eta^\prime$
mesons from the primary ${\gamma}N\to {\eta^\prime}N$ channel in the interaction of photons of energies
of 1.5--2.2 GeV with $^{12}$C (left panel) and $^{93}$Nb (right panel) nuclei. For notation see the text.}
\label{void}
\end{center}
\end{figure}

  In addition, we have investigated the momentum dependence of the absolute and relative
$\eta^\prime$ meson yields for incident photon energies between 1.5 and 2.2 GeV. The momentum
differential cross sections for $\eta^\prime$ production from $^{12}$C and $^{93}$Nb, calculated
on the basis of Eq. (9) in the scenario of collisional broadening of the $\eta^\prime$ meson
characterized by the value of $\sigma_{{\eta^\prime}N}=8$ mb extracted above with an in-medium
${\eta^\prime}$ mass shift (6) (dashed curve) and without it (solid curve), are depicted in
figure 7.
It is seen that in the case, when an in-medium ${\eta^\prime}$ mass shift is included, their production
cross sections increase as compared to those obtained in the scenario without it
at ${\eta^\prime}$ momenta $\le$ 1 GeV/c for both target nuclei.
This gives an opportunity to get some information on a possible
${\eta^\prime}$ mass shift in nuclear matter from a precise measurement of the momentum distribution.
Such possibility has been also discussed recently for the $\omega$ meson in [36].

    The experimental data [19] on the momentum distribution of $\eta^\prime$ mesons produced off a
$^{12}$C target for $E_{\gamma}=$ 1.5--2.2 GeV are shown in figure 8 in comparison to the calculated
momentum distribution given in figure 7 and normalized to the data point at $\eta^\prime$ momentum of
1.1 GeV/c. One can see that the available data seem do not exclude both the scenario with and the scenario
without $\eta^\prime$ mass shift, which means that the absolute experimental cross sections have to be
determined to shed light on this shift.
\begin{figure}[!h]
\begin{center}
\includegraphics[width=10.0cm]{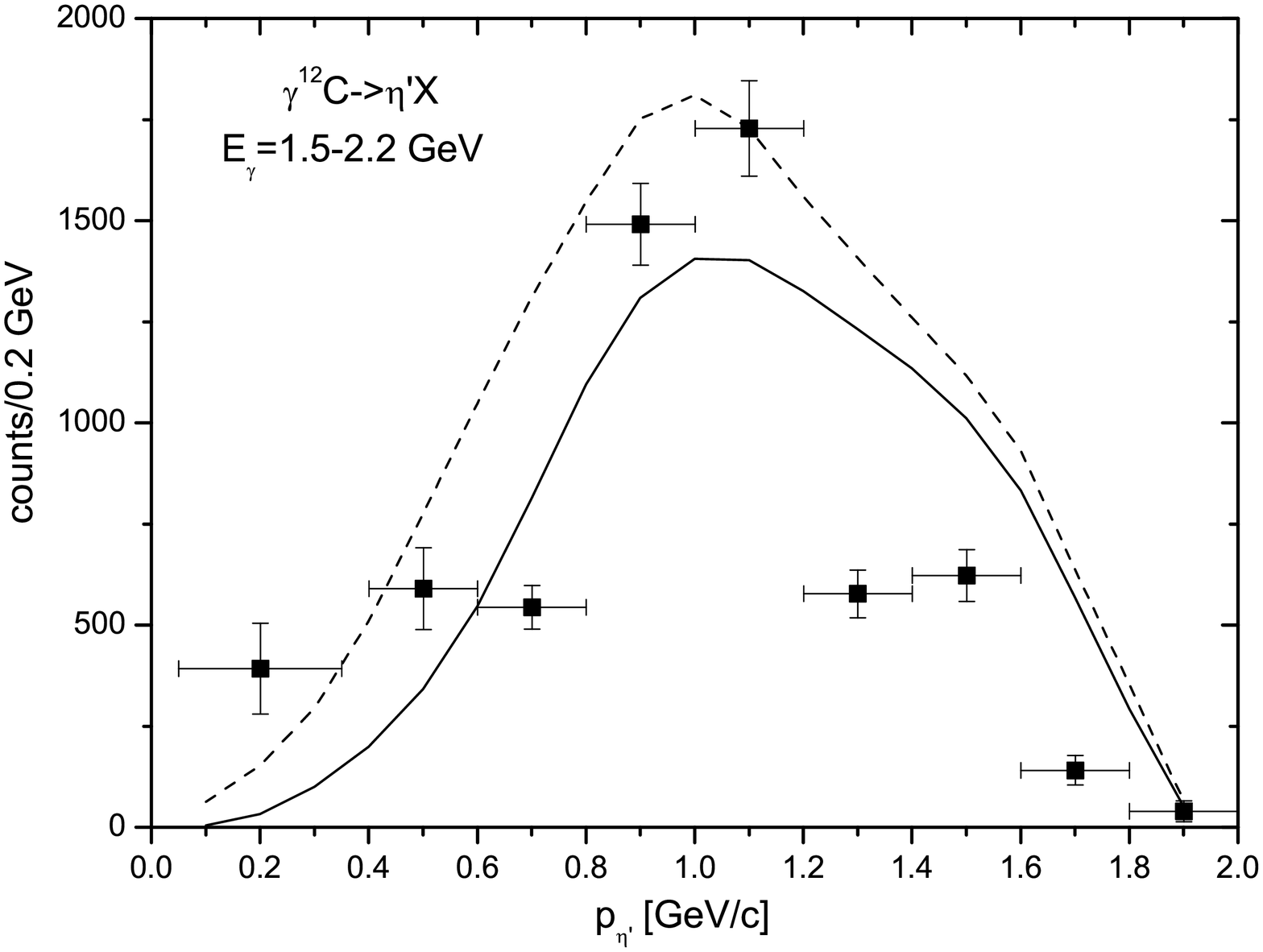}
\vspace*{-2mm} \caption{Momentum distribution of $\eta^\prime$
mesons produced off a $^{12}$C target for $E_{\gamma}=$ 1.5--2.2 GeV.
The experimental data are from [19]. The curves are our
calculations normalized to the data point at $\eta^\prime$ momentum of 1.1 GeV/c.
The notation of the curves is identical to that in figure 7.}
\label{void}
\end{center}
\end{figure}
\begin{figure}[!b]
\begin{center}
\includegraphics[width=12.0cm]{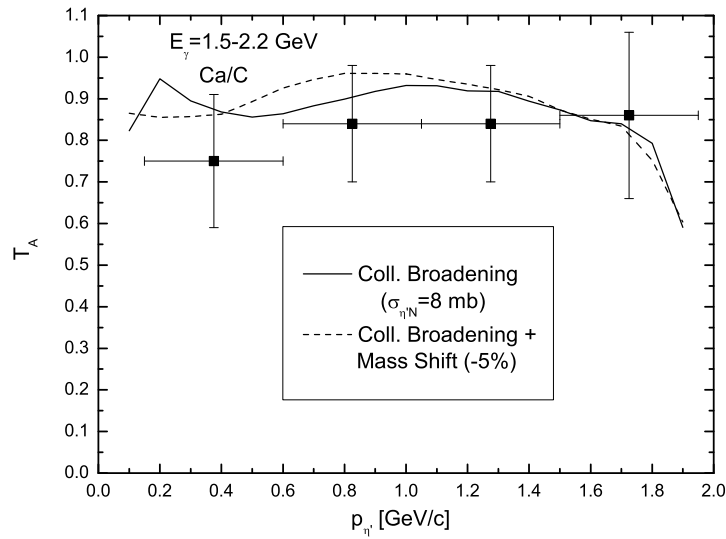}
\vspace*{-2mm} \caption{Transparency ratio $T_A$ as a function of the $\eta^\prime$ momentum
for combination Ca/C and for $E_{\gamma}=$ 1.5--2.2 GeV.
For notation see the text.}
\label{void}
\end{center}
\end{figure}
\begin{figure}[!h]
\begin{center}
\includegraphics[width=12.0cm]{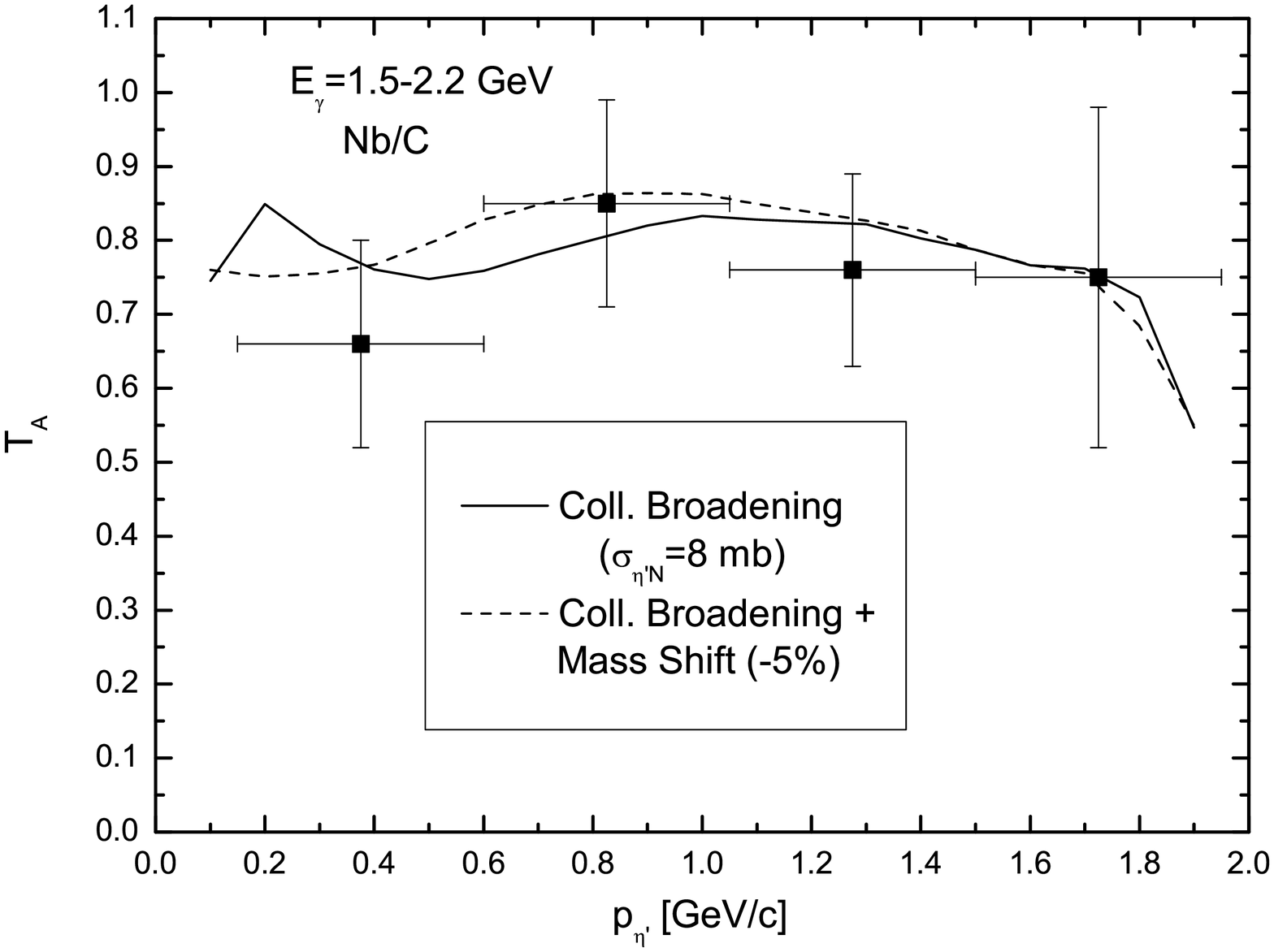}
\vspace*{-2mm} \caption{The same as in figure 9, but for the combination Nb/C.}
\label{void}
\end{center}
\end{figure}
\begin{figure}[!h]
\begin{center}
\includegraphics[width=12.0cm]{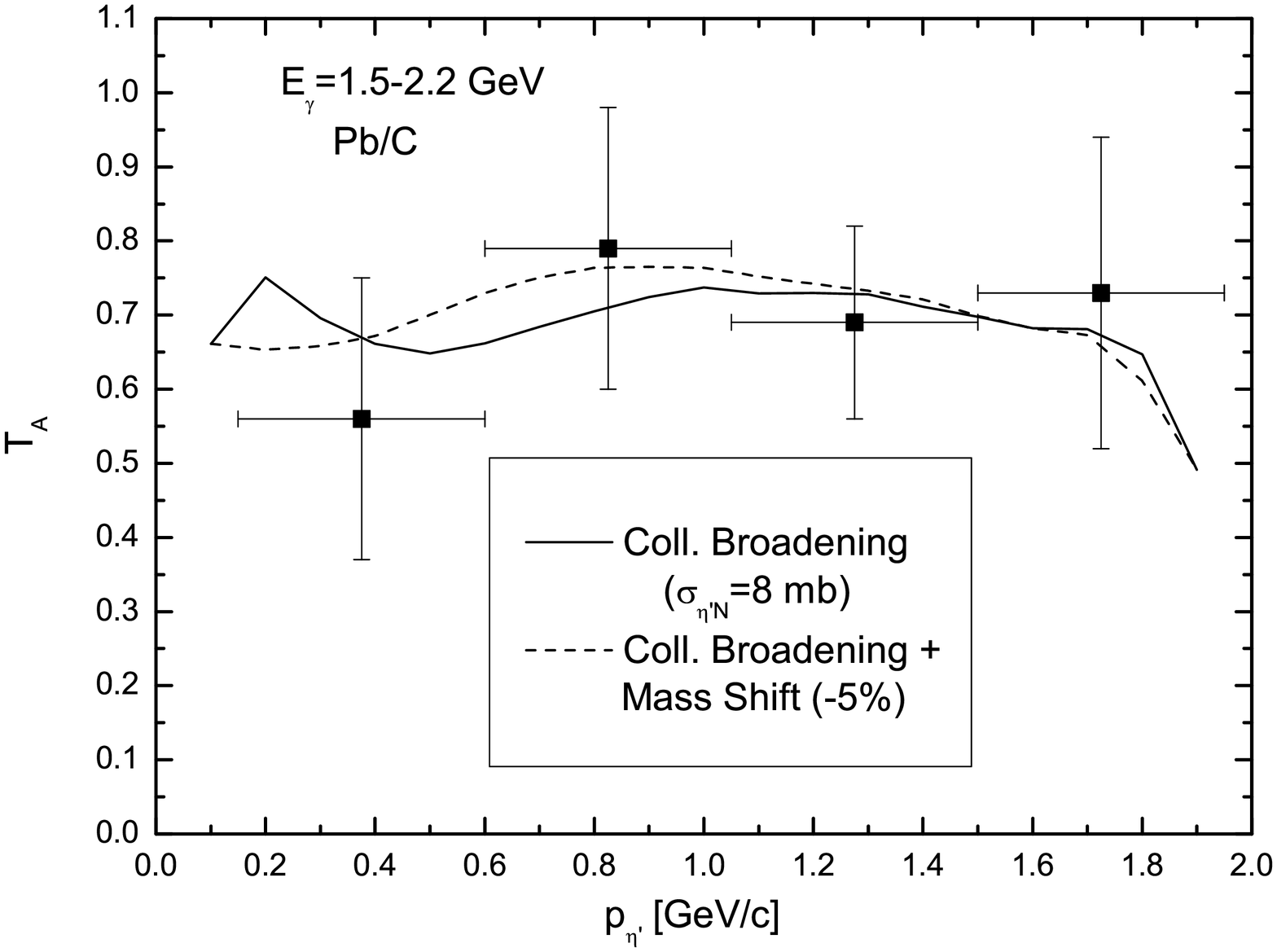}
\vspace*{-2mm} \caption{The same as in figure 9, but for the combination Pb/C.}
\label{void}
\end{center}
\end{figure}
\begin{figure}[!h]
\begin{center}
\includegraphics[width=12.0cm]{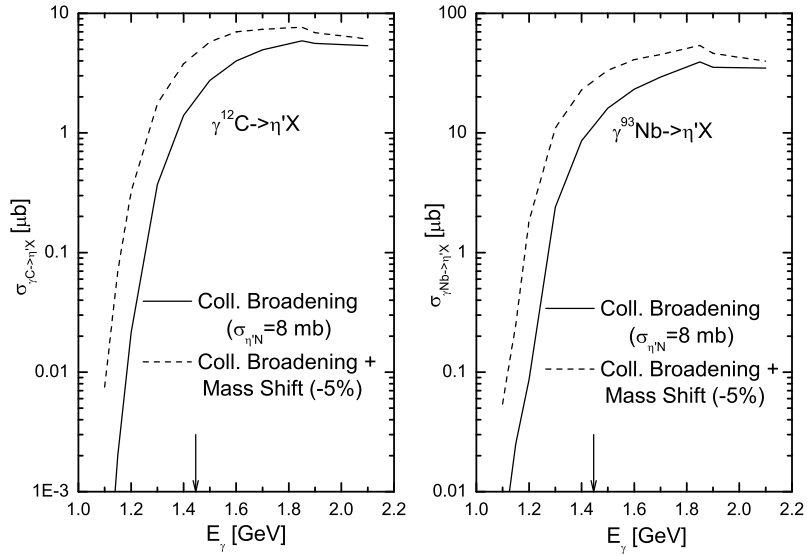}
\vspace*{-2mm} \caption{Excitation functions for photoproduction of $\eta^\prime$
mesons off $^{12}$C (left panel) and $^{93}$Nb (right panel), respectively. The dashed and solid
curves are calculations, assuming collisional broadening of the $\eta^\prime$ meson with an in-medium
$\eta^\prime$ mass shift (6) and without it, correspondingly. The arrows indicate the threshold energy
for $\eta^\prime$ photoproduction on a free nucleon.}
\label{void}
\end{center}
\end{figure}

  The measured (full squares) in [19] momentum dependence of the transparency ratio $T_A$ for
$\eta^\prime$ mesons for combinations Ca/C, Nb/C and Pb/C and for the photon energy range of
1.5--2.2 GeV is shown, respectively, in figures 9, 10 and 11 in comparison to our calculations
on the basis of Eq. (9), assuming collisional broadening of the $\eta^\prime$ meson
characterized by the value of $\sigma_{{\eta^\prime}N}=8$ mb extracted above from the analysis
of the $A$--dependence of this ratio with an in-medium
${\eta^\prime}$ mass shift (6) (dashed curve) and without it (solid curve).
It is clearly seen that, on the one hand, the differences between the calculations corresponding to
different assumptions about the ${\eta^\prime}$ mass shift (between dashed and solid curves) are
insignificant for all considered $A$/C combinations and, on the other hand, both these calculations
describe quite well the data [19]. We, therefore, come to the conclusion that the  transparency ratio
measurements [19] do not allow one to discriminate between two adopted scenarios for the in-medium
${\eta^\prime}$ modification.

    Finally, we consider the excitation functions for photoproduction of $\eta^\prime$ mesons off
$^{12}$C and $^{93}$Nb target nuclei. They were calculated for the one-step $\eta^\prime$ production
mechanism on the basis of Eq. (10) as well as for $\sigma_{{\eta^\prime}N}=8$ mb and for two adopted
options for the $\eta^\prime$ in-medium mass shift, and are given in figure 12. Looking at this figure,
one can see that there are well separated predictions for
this two considered scenarios for the $\eta^\prime$ in-medium
mass shift. Therefore, one can conclude that the excitation function measurements might help to get a
valuable information about this shift. It should be mentioned that an analogous possibility has been
discussed before for the $\omega$ meson in [36].

   Taking into account the above considerations, we come to the conclusion that such observables as
the momentum distribution and excitation function for $\eta^\prime$ photoproduction from nuclei
can be useful to help determine a possible $\eta^\prime$ mass shift in cold nuclear matter.

\section*{4. Conclusions}

\hspace{1.5cm} In this paper we have investigated the production of $\eta^\prime$ mesons
in photon-induced nuclear reactions near the threshold on the basis of an approach,
which accounts for both
primary photon--nucleon and secondary pion--nucleon $\eta^\prime$ production processes
as well as two different scenarios for the $\eta^\prime$ in-medium mass shift.
We have found that the secondary channel ${\pi}N \to {\eta^\prime}N$ plays a minor role in
$\eta^\prime$ photoproduction off nuclei. Hence, it can be ignored in extracting the inelastic
${\eta^\prime}N$ cross section $\sigma_{\rm {\eta^\prime}N}$ from the analysis within the
present approach of the available data [19] on the $\eta^\prime$ transparency ratio. We have
deduced from this analysis that $\sigma_{\rm {\eta^\prime}N}\approx$ 6--10 mb. This value
coincides with that obtained in [19] in a low-density approximation. We have shown
that the transparency ratio measurements [19] do not allow one to discriminate between two adopted
scenarios for the $\eta^\prime$ in-medium mass shift. On the other hand, we have also shown that,
contrary to the transparency ratio, both the momentum distribution and excitation function for
$\eta^\prime$ photoproduction off nuclei are appreciably sensitive to the shift. This gives an
opportunity to determine it experimentally.
\\
\\
{\bf Acknowledgments}
\\
\\
The author gratefully acknowledges V. Metag and M. Nanova for their information
on experimental results on $\eta^\prime$ meson photoproduction off nuclei as well as for their
interest in this work.
\\
\\


\begin{thebibliography}{99}
\bibitem{1} T. Ishikawa {\it et al.}, Phys. Lett. B {\bf 608}, 215 (2005).
\bibitem{2} C. Djalali {\it et al.}, J. Phys. G. {\bf 35}, 104035 (2008).
\bibitem{3} M. H. Wood {\it et al.}, Phys. Rev. C {\bf 78}, 015201 (2008).
\bibitem{4} M. H. Wood {\it et al.}, Phys. Rev. Lett. {\bf 105}, 112301 (2010).
\bibitem{5} M. Kotulla {\it et al.}, Phys. Rev. Lett. {\bf 100}, 192302 (2008).
\bibitem{6} M. Nanova {\it et al.},  Phys. Rev. C {\bf 82}, 035209 (2010).
\bibitem{7} M. Nanova {\it et al.},  Eur. Phys. J. A {\bf 47}, 16 (2011).
\bibitem{8} M. Naruki {\it et al.}, Phys. Rev. Lett. {\bf 96}, 092301 (2006).
\bibitem{9} R. Muto {\it et al.}, Phys. Rev. Lett. {\bf 98}, 042501 (2007).
\bibitem{10} A. Polyanskiy {\it et al.}, Phys. Lett. B {\bf 695}, 74 (2011).
\bibitem{11} M. Hartmann {\it et al.}, Phys. Rev. C {\bf 85}, 035206 (2012).
\bibitem{12} G. Agakishiev {\it et al.}, Phys. Rev. C {\bf 84}, 014902 (2011).
\bibitem{13} D. Adamova {\it et al.}, Phys. Rev. Lett. {\bf 91}, 042301 (2003);\\
             S. Damjanovic {\it et al.}, Eur. Phys. J. C {\bf 49}, 235 (2007);\\
             R. Arnaldi {\it et al.}, Eur. Phys. J. C {\bf 61}, 711 (2009).
\bibitem{14} H. Nagahiro, M. Takizawa, and S. Hirenzaki, Phys. Rev. C {\bf 74}, 045203 (2006).
\bibitem{15} D. Jido, H. Nagahiro, and S. Hirenzaki, Phys. Rev. C {\bf 85}, 032201 (2012).
\bibitem{16} T. Cs$\ddot{\rm o}$rgo, R. Vertesi, and J. Sziklai, Phys. Rev. Lett. {\bf 105}, 182301 (2010).
\bibitem{17} P. Moskal {\it et al.}, Phys. Lett. B {\bf 474}, 416 (2000);\\
             P. Moskal {\it et al.}, Phys. Lett. B {\bf 482}, 356 (2000).
\bibitem{18} E. Oset and A. Ramos, Phys. Lett. B {\bf 704}, 334 (2011).
\bibitem{19} M. Nanova {\it et al.}, Phys. Lett. B {\bf 710}, 600 (2012).
\bibitem{20} K. Itahashi {\it et al.}, arXiv: 1203.6720 [nucl-ex].
\bibitem{21} S. V. Efremov and  E. Ya. Paryev, Eur. Phys. J. A {\bf 1}, 99 (1998).
\bibitem{22} E. Ya. Paryev, Eur. Phys. J. A {\bf 7}, 127 (2000).
\bibitem{23} E. Ya. Paryev, J. Phys. G. {\bf 36}, 015103 (2009).
\bibitem{24} E. Ya. Paryev, arXiv: 1112.0153 [nucl-th].
\bibitem{25} I. Jaegle {\it et al.}, Eur. Phys. J. A {\bf 47}, 11 (2011).
\bibitem{26} H. Nagahiro {\it et al.}, Phys. Lett. B {\bf 709}, 87 (2012).
\bibitem{27} V. K. Magas {\it et al.}, arXiv: 0911.3614 [nucl-th].
\bibitem{28} Ye. S. Golubeva {\it et al.}, Nucl. Phys. A {\bf 625}, 832 (1997).
\bibitem{29} M. Dugger {\it et al.}, Phys. Rev. Lett. {\bf 96}, 169905 (2006).
\bibitem{30} M. Williams {\it et al.}, Phys. Rev. C {\bf 80}, 045213 (2009).
\bibitem{31} V. Crede {\it et al.}, Phys. Rev. C {\bf 80}, 055202 (2009).
\bibitem{32} E. Ya. Paryev, Eur. Phys. J. A {\bf 23}, 453 (2005).
\bibitem{33} E. Ya. Paryev, Eur. Phys. J. A {\bf 5}, 307 (1999).
\bibitem{34} P. M$\ddot{\rm u}$hlich and U. Mosel, Nucl. Phys. A {\bf 773}, 156 (2006).
\bibitem{35} V. Flaminio {\it et al.}. Compilation of cross sections.\\
             I: $\pi^+$ and $\pi^-$ induced reactions. CERN--HERA {\bf 83--01} (1983).
\bibitem{36} V. Metag {\it et al.}, Prog. in Part. and Nucl. Phys. {\bf 67}, 530 (2012).
\end{thebibliography}
\end{document}